# Division of the Energy Market into Zones in Variable Weather Conditions using Locational Marginal Prices

Karol Wawrzyniak, Grzegorz Oryńczak, Michał Kłos, Aneta Goska, Marcin Jakubek
National Centre for Nuclear Research, Świerk Computing Centre
Otwock-Świerk, Poland
K.Wawrzyniak@fuw.edu.pl

*Abstract*— Adopting a zonal structure of electricity market requires specification of zones' borders. One of the approaches to identify zones is based on clustering of Locational Marginal Prices (LMP). The purpose of the paper is twofold: (i) we extend the LMP methodology by taking into account variable weather conditions and (ii) we point out some weaknesses of the method and suggest their potential solutions. The offered extension comprises simulations based on the Optimal Power Flow (OPF) algorithm and twofold clustering method. First, LMP are calculated by OPF for each of scenario representing different weather conditions. Second, hierarchical clustering based on Ward's criterion is used on each realization of the prices separately. Then, another clustering method, i.e. consensus clustering, is used to aggregate the results from all simulations and to find the global division into zones. The offered method of aggregation is not limited only to LMP methodology and is universal.

## I. INTRODUCTION

The whole energy market of Europe is under an intensive process of transformation. The main drivers for change are integration of markets and growing use of renewable generation. Currently, the most popular market structures are uniform, nodal, and zonal pricing. The former is still used in many countries mainly due to historical reasons. In spite of its apparent simplicity, such an approach has serious disadvantages. The equilibrium set on the market does not take into account safety requirements of the grid. Hence, (i) the single-price equilibrium set on the market (energy exchange) is frequently unfeasible, (ii) the system operator has to perform costly readjustments, (iii) costs of supplying the energy differ between locations, but are not covered where they arise[1]. Introducing other forms of market helps to eliminate congestion costs. Hitherto, the explicit type of the future pan-European energy market remains an open question as the Third Energy Package, especially regulations 713/2009, 714/2009 and directive 2009/72/WE, does not specify it precisely.

Wholesale electricity markets use different market designs to handle congestion in the transmission network. The two most popular approaches towards which national markets evolve are nodal and zonal pricing. The nodal pricing model is currently used in, among others, the US and Russia. Zonal pricing has been introduced in the Nordic countries as well as in Great Britain. It uses the market coupling algorithm to calculate prices in zones given the possible transfers between them. Considering dividing the market into zones it is worth noticing that over the last few years there has been a steady rise in the volume of cross border trade, however, it was not met by a considerable growth in the cross border transmission capacities. Hence, country borders with their strict congestion limits seem to be natural candidates for zone borders.[2] In consequence, the questions which arise are whether and how a national market should be divided into zones. The existing methods are mostly based on two-stage approach – assignment of specific values to each of the nodes and division of the system into regions by clustering the nodes. Among existing methods for completing the first stage we can point out two concurrent ways of reasoning.

The first aims at calculating Power Transfer Distribution Factors (PTDFs) and using them to aggregate nodes [1],[2]. The assignment depends only on the parameters that refer to congested lines. The distribution factors reflect the influence of unit nodal injections on power flow along the transmission lines, thus grouping the nodes characterized by similar factors into one zone defines a region of desirably similar sensitivity to congestions.

The second approach is based on *nodal prices* called also Locational Marginal Prices (LMP) [3],[4]. Nodal pricing is a method of determining prices in all locations of the transmission grid. These locations – nodes – are representing physical points of the transmission system, where the energy can be injected by generators or withdrawn by loads. The price at each node represents the locational value of energy i.e. a cost of supplying extra 1 MW of energy to node. It consists of the cost of energy used at a node and the cost of delivering it there. The latter depends on losses and congestion. This approach utilizes the fact that each congestion leads to graduate increase of delivery cost [5]. Thus, aggregation of similar nodal prices should result in a reliable solution.

---

[1] For example, in Poland in 2011 the cost of the balancing market readjustments amounted to more than 3% (>250Mln EUR) of the overall costs of production (*source: URE/ARE S.A.*).

This work was supported by the EU and MSHE grant nr POIG.02.03.00-00-013/09. The use of the CIS computer cluster at NCBJ is gratefully acknowledged.

[2] Taking as an example the Nordic countries, one can find that zones are created within the borders of each country.

However, in the literature concerning division into zones [3],[4] usually stable levels of generation are assumed, which remains in contradiction with the increasing amount of renewable generation for which, as yet, wind farms, characterized by highly variable power output, constitute the main source. We found that the relative instability in the amount of power injected into the system by wind farms significantly influences the energy prices even if the rate of wind generation to total generation is relatively small. Hence, we extend the LMP method by taking into account variable weather conditions. In essence, we postulate adding a third, "aggregating" step to the method, after the calculation of LMP and clustering was conducted for each weather scenario. We also present some critical remarks related to results that the method can lead to.

The exact methodology is presented in Sec. II. Then, in Sec. III the results of the research, including limitations of the method, are presented and discussed. Finally, in Sec. IV, we point out issues which would be necessary to make the study more conclusive.

## II. THE METHODOLOGY

The methodology used in this paper to determine division of energy market into zones is based on the concept of nodal pricing. The problem of finding nodal prices is solved using Direct Current version of OPF. The objective of DC OPF is to find a steady state operation point which minimizes generation cost, which is identical with maximization of social welfare (understood in this case as consumer surplus due to inflexibility of demand profile) under the constraints of the transmission system and assumption that losses can be neglected.[3] The areas where congestion does not exist are characterized by similar nodal prices. Such groups are suitable candidates for zones in the zonal pricing model. Hence, using clustering algorithms, one can group nodes of similar prices [3],[4],[6]. The task is relatively easy assuming stability of power generation. In reality, the load, the generation and, in turn, prices, fluctuate, and can influence the division into zones. Hence, a robust approach has to take into account various circumstances and aggregate divisions obtained for different cases in order to produce a "generalized" division. We do this using $M = 722$ historical wind scenarios leading to different wind farms generations.[4] OPF is then run to determine nodal prices. Next, a clustering algorithm is used to produce a zonal division for each realization of nodal prices. Then, the results of clusterings are aggregated using another clustering technique ("consensus clustering") leading to optimal division. The general scheme of the research (Fig. 1) is discussed in detail in the next subsections.

---

[3] Neglecting the losses (a fundament of DC approximation) allows us to concentrate on the variation in nodal prices which arises only as a result of congestion cost.

[4] Other possibility would be to use Monte Carlo (MC) technique, where wind realizations are drawn randomly according to probability density functions estimated for wind farm locations.

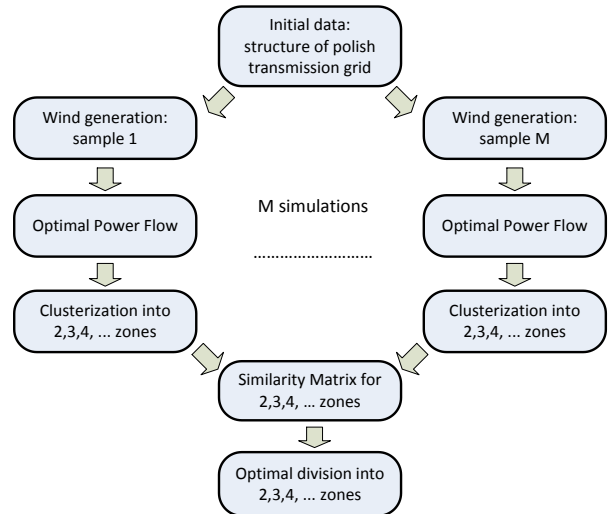

Figure 1. Scheme of the research.

### A. Data description

In our research the Polish power grid is analyzed as an example. The model has been based on exemplary case file included in MATPOWER distribution [7]. This case represents the Polish 400, 220 and 110 kV network during winter 1999-2000 peak conditions. The system consists of 327 power generators, 2383 buses, and 2896 interconnecting branches. All lines included in the profile are of high (110 kV) or extra high (220 kV and 400 kV) voltage levels. The quality of the data is unknown but analysis of the assumed costs of generation gives the premise that these data are rather of poor quality. We used this case instead of other cases related to more recent periods due to fact that this the only one where congestions exist under base case load.

### B. Wind generation

Considering wind generation, in 2000 wind farms in Poland hardly existed. For sake of simplicity, we added the current wind generation to the grid existing in 2000 by inserting new buses and branches connecting the generators wind farms to the grid.

First, using information provided by The Polish Wind Energy Association we localized the nodes of wind farm grids to be added. We focused only on the biggest farms, generating not less than 5 MW each. As a result, we obtained 39 new nodes distributed all over the country with generation impact of a single source reaching 120 MW. Cumulative power of the farms that we identified is 1.4 GW. This constitutes 64% of overall wind generation in Poland in 2012. Since we did not possess data about the real location and parameters of the branches, we created them artificially in such a way that the bus of a wind generator is connected to the closest 220 kV or 400 kV bus by a standardized branch. The profile was used to construct 39 new transmission lines. We are aware that such an approach does not fully reflect the complex reality, where branches are differently characterized and can be potentially congested. Considering both the quality of original data and the way how the wind generation is added, the data should be treated as an exemplary case useful only for testing the method but not to draw any kind of valuable conclusions related to the Polish electric system.

As the wind is not steady, it is necessary to know the probability density function of the wind speed across the country or to have at the disposal historical data related to farms' locations. In our research we used data from the National Climate Data Center [8]. The data includes wind measurement from 139 weather stations located in the territory of Poland. We analyzed data for winter months (Nov - Feb) and years 2007 - 2012. In this way we obtained 722 different scenarios of historical wind speeds in different locations which we then used for simulations where one scenario corresponds to one day. For every wind farm the nearest weather station (in the sense of Euclidean distance) is assigned. The wind speed $v_w^0$ of $w = 1,...,W$ farm is determined by the weather station at height $H_w^0 = 10$ meters above ground level. Since the wind $v_w$ on the turbine level $H_w$ is higher than on $H_w^0$, we extrapolated the data to the height of wind turbines using the one seventh power law [9]: $v_w = v_w^0 \left( H_w / H_w^0 \right)^\alpha$, where $v_w$ - mean wind speed at farm $w$ which turbine is located at height $H_w$. We assumed $H_w = H = 80\,\text{m}$. The coefficient α is the friction coefficient (Hellman exponent). This coefficient is a function of the topography at a specific site and frequently assumed as a value of 1/7 for open land.

Using a sigmoid function which estimates relation between wind speed in the area and power generated by a farm located there, we calculated the power generated by each of 39 wind farms that we identified. The function estimated for Vestas V112 3.0 MW turbine [10] was assumed as the common output profile for all farms: $O_w = O_w^{max} / (1 + 439 e^{-0.8 v_w})$, where $O_w$ is the output of wind farm $w$ in MW, and $O_w^{max}$ describes maximal possible output of the wind farm $w$.

### C. Optimal Power Flow

LMPs are obtained by running Direct Current OPF in MATPOWER simulator [7]. The algorithm finds levels of generations which minimize the total cost of system operation under limits on energy network transmission capabilities. Using DC version of OPF instead of Alternating Current (AC) version is dictated by the fact that losses are then omitted and the only premise to create zones is to avoid persistent congestions within zones. This problem is widely discussed in the literature, therefore we do not provide here more detailed description.

### D. Clustering of nodal prices

Having obtained the nodal prices $\left( P_n^m \right)$ for each of $n = 1,...,N$ nodes in $m$-th, $m = 1,...,M$, scenario of the wind generators' outputs, we proceed to determine the (first-fold)[5] division of the country-wide energy network into zones, which are defined as connected sets of nodes with "similar" nodal prices of energy. In order to group nodes into zones, we use hierarchical agglomerative clustering of the nodal prices based on Ward's criterion [11], modified to acknowledge the existence of connection (branch) between nodes of the clusters.

This clustering technique has two characteristics which are especially useful in our study. First, the number of clusters into which the set of observations is to be divided may be not known *ex ante*: the obtained hierarchy of clusters' mergers allows to divide the set of observations into any number of clusters, examine characteristics of each division and decide on the number of clusters *ex post*. Second, this clustering method can be straightforwardly adapted to recognize the topology of the graph representing the energy network in order to merge only those clusters which have a connection between them (that is, there exists a branch connecting a node from one cluster with a node from the other cluster).

As the (dis)similarity measure determining optimal clusters to be merged, we use Ward's criterion [11], which is of a special appeal to our study, since it often generates clusters of similar sizes [12] - a feature which is desired for deriving market zones of substantial and comparable scales. To incorporate the topology of the network, during the computations we augment the dissimilarity measure so that any not connected clusters have dissimilarity of $+\infty$.

Specifically, to review step by step the process described above: for a given realization $m \in \{1,...,M\}$ of nodal prices computation, we start with $N$ singleton-clusters $C_1,...,C_N$ representing each node in the network, described by its nodal price, $P_n$, as the single feature characterizing it.[6] The Ward's criterion seeks to minimize the total inner-cluster error sum of squares (ESS), which for division into $K$ clusters $C_1,...,C_K$ amounts to $ESS_K = \sum_{k=1}^{K} E(C_k)$, where

$$E(C_k) = \sum_{n \in C_k} \left( P_n - \overline{P}(C_k) \right)^2,$$

---
[5] "First-fold," since the divisions obtained for each of *M* wind scenarios will be then aggregated to obtain a concluding market zones (cf. Sec. II.E).
[6] Since we discuss below clustering of only one realization, we omit the superscript *m* of the nodal price for the sake of readability.

$$\overline{P}(C_k) = \frac{1}{|C_k|} \sum_{n \in C_k} P_n$$ and $|C_k|$ is the cluster-$k$ size.

In the first step of the algorithm, from all the singleton-clusters $C_1, ..., C_N$ two clusters, $C_k, C_l$, $k,l \in \{1,...,N\}, k \neq l$ are chosen to merge. The resulting cluster $C_k \cup C_l$ yields minimal increase in the total inner-cluster ESS, taking into account the connectivity of the branches, that is,[7]

$$\{k,l\} = \arg\min_{k,l} \{\tilde{E}(C_k \cup C_l) - E(C_k) - E(C_l)\},$$

where

$$\tilde{E}(C_k \cup C_l) = \begin{cases} E(C_k \cup C_l) & \text{if there exists branch } b(i,j) \\ & \text{for some } i \in C_k, j \in C_l, \\ +\infty & \text{otherwise.} \end{cases}$$

The clusters $C_k, C_l$ are then removed from the candidates to merge, and the new cluster $C_k \cup C_l$ enters the candidates. The above step is then repeated $N-1$ times up to the point when all the observations form a single cluster. Tracing back the mergers (starting from the last), we can then derive division of the entire set of nodes into any number of $2, ..., N$ clusters (viz. market zones).

The co-occurrences of nodes in divisions into 2, 3, 4 clusters for each of the $M$ wind realizations form then a basis for an aggregated division into market zones, which is discussed in the next section.

*E. Similarity matrix & optimal division*

Obviously, the assignment of nodes into zones can vary from one wind scenario to another. Finally, the results have to be aggregated in one structure which reflects how frequently every pair of nodes belongs to the same cluster. This problem is known as *consensus clustering* or aggregation of clustering. It refers to the situation in which a number of different clustering results have been obtained from different runs of the same clustering method. The task of the algorithm is to create a single (consensus) clustering which generalizes the results of a whole set of runs. In our approach we use Cluster-based Similarity Partitioning Algorithm (CSP) [13]. Essentially, if two objects are in the same cluster then they are considered to be fully similar, and if not they are marked dissimilar. Similarity between two objects takes the value of 1 if they are in the same cluster and 0 otherwise. For each clustering, i.e. one scenario, a binary similarity matrix is created. The entry-wise average of such matrices representing the sets of groupings yields an overall *similarity matrix*. Finally we normalize values to estimate the probability that two nodes are connected. The formal notation is as follows.

Again, let $N$ be the number of nodes and $M$ be the number of wind scenarios. For each scenario $m$, $m \in \{1,...,M\}$, we calculate binary similarity matrix $S_m$, where $S_m \in \mathbb{B}^{N \times N}, \mathbb{B} = \{0,1\}$. Then, the overall similarity matrix is created with elements

$$S(i,j) = \frac{1}{M} \sum_{m=1}^{M} S_m(i,j).$$

Next, the normalized similarity matrix is used to recluster the objects. We developed a sequential technique where in each step two clusters or nodes characterized by the highest similarity measure (value in similarity matrix) are merged together.

In the first step of the algorithm, $t_1$, the similarity matrix is duplicated $S^{t_1} := S$, expressing similarity between the singleton-clusters from set $Z^{t_1} = \{Z_1^{t_1}, ..., Z_N^{t_1}\}$. Then, in every step $t \in \{t_1, t_2, ..., t_{N-1}\}$, the most similar two singleton clusters $Z_k^t, Z_l^t$ are merged into a new cluster $Z_k^t \cup Z_l^t$, and the new set of clusters, $Z^{t+1}$, is created: $Z^{t+1} = \{Z_k^t \cup Z_l^t\} \cup \{Z^t \setminus \{Z_k^t, Z_l^t\}\}$.

The similarity measure for pairs of elements in newly created set $Z^t$ is recalculated based on the genuine matrix $S$, according to:[8]

$$S^t(k,l) = \frac{1}{|Z_k^t||Z_l^t|} \sum_{i \in Z_k^t} \sum_{i \in Z_l^t} S(i,j)$$

Obviously, if nodes $i$ and $j$ are always in the same cluster then $S(i,j) = 1$. If not, $S(i,j)$ reflects the frequency with which nodes $i$ and $j$ were in the same clusters across the wind data sample. Hence, the dimension of matrix $S^t$ is reduced by one in each step, $S^t \in \mathbb{R}^{(N-t+1) \times (N-t+1)}$. This procedure is then repeated and stops when a desired number of clusters is achieved.

III. RESULTS & DISCUSSION

In our analysis, the OPF run across the 722 wind scenarios identified 9 different branches in the network model where congestion could arise. The exact number of congested branches varies from one wind scenario to another. The different locations and levels of congestion across scenarios are reflected in different values of nodal prices. In the two extreme wind scenarios, only 70% of buses are assigned to analogous clusters. These cases are characterized by the average wind speed at 80 m height level equal to 12.4 m/s and 24.2 m/s, respectively. Comparison of the cases provides clear evidence that wind generation has to be taken into account when division into zones is considered.

---

[7] Of course, when merging the singleton clusters, we have $E(C_k) = E(C_l) = 0$. However, in the succeeding steps of the algorithm, the merging clusters may have non-zero inner ESS.

[8] When a real implementation is considered there is no need to recalculate similarity measure for all pairs in set $Z^t$. It is enough to recalculate only values for newly created element $Z_k^t$ and any other element $Z_l^t$.

For choosing the exact number of zones to which market should be divided there is, as yet, no widely accepted methodology, although there are some attempts [6]. In our study we assume the range of divisions to, respectively, 2, 3, and 4 clusters and we analyze each of them separately.

The results for LMP division are in Tab. 1 and Fig. 2. In most cases some of the resulting clusters are extremely tiny. It happens if small regions are isolated by congested lines. Usually these tiny clusters comprise only few loads and no generation. Tiny clusters can be hard to accept as a separate zone for a few reasons. The zone should define its own "energy market," therefore, generation and load in a zone should, at least partially, balance itself. Otherwise, we end up with a zone which is purely an importer of energy, and the only solution to providing necessary amount of power to this zone is based on readjustments made by the TSO in neighboring zones, which is inconsistent with the idea of zonal approach to energy markets. Another obstacle is rather sociological. It may be difficult for the society to accept a tiny zone where price is manifold higher than in neighboring areas. In literature we have not found the methodology explaining how to treat small zones. We suggest integrating such zone with a neighboring zone for which the Ward's criterion is the lowest. The critical size of the cluster, being a lower limit from which a cluster should be considered as a zone, remains an open question.

Another issue that requires manual interference is related to clusters intermingling (see Fig. 2 and Tab. 1). The reason behind the intermingling is usually related to poor integration of networks of different voltages. E.g. in Fig. 2 (bottom, left) 400 kV line are above 220 kV lines and there is no common crossing point in neighboring area. Intermingling causes that a geographical location does not determine explicitly the zone to which the node is assigned. This can lead to lower transparency of zonal markets where zones are not geographically consistent. In currently used structures, like Nordpool, the geographical integration is required. Again, the known literature does not state how to proceed with intermingling cases. Manual identification of intermingling area, its separation, measurement of Ward's criterion and reconsidering an aggregation with the most suitable neighbor, can be a potential solution. Considering the properties of other clustering methods, like Fuzzy C-mins [14] or Price's Differential Clustering [4], can potentially lead to development of the method where intermingling is avoided.

| | | # nodes | total power demand [GW] | # generators | total output [GW] | marginal price [PLN/MWh] | average nodal price [PLN/MWh] |
|---|---|---|---|---|---|---|---|
| 2 clusters | | 2325 | 23.0 | 357 | 24.1 | 170.7 | 145 ± 22 |
| | | 2400 | 24.3 | 366 | 24.6 | 170.7 | 145 ± 27 |
| | | 2401 | 24.3 | 366 | 24.6 | 170.7 | 144 ± 21 |
| | | 97 | 1.6 | 9 | 0.5 | 0 | 320 ± 114 |
| | | 21 | 0.3 | 0 | 0 | 0 | 422 ± 69 |
| | | 22 | 0.3 | 0 | 0 | 0 | 353 ± 47 |
| 3 clusters | yellow | 2325 | 23.0 | 357 | 24.1 | 170.7 | 145 ± 22 |
| | | 2234 | 21.8 | 343 | 23.3 | 170.7 | 140 ± 19 |
| | | 1592 | 15.3 | 218 | 18.9 | 165.6 | 134 ± 8 |
| | green | 22 | 0.3 | 0 | 0 | 0 | 505 ± 89 |
| | | 22 | 0.3 | 0 | 0 | 0 | 422 ± 69 |
| | | 21 | 0.3 | 0 | 0 | 0 | 353 ± 47 |
| | purple | 75 | 1.3 | 9 | 0.5 | 0 | 266 ± 39 |
| | | 166 | 2.5 | 23 | 1.3 | 142.9 | 209 ± 34 |
| | | 809 | 8.9 | 148 | 5.6 | 170.7 | 163 ± 24 |
| 4 clusters | yellow | 1477 | 14.8 | 255 | 12.1 | 170.7 | 154 ± 20 |
| | | 1286 | 12.6 | 163 | 15.5 | 165.6 | 132 ± 18 |
| | | 1592 | 15.3 | 218 | 18.9 | 165.6 | 134 ± 8 |
| | grey | 848 | 8.1 | 102 | 11.9 | 152.9 | 128 ± 11 |
| | | 948 | 9.2 | 180 | 7.8 | 170.7 | 151 ± 16 |
| | | 734 | 7.6 | 139 | 5.1 | 170.7 | 158 ± 15 |
| | purple | 75 | 1.3 | 9 | 0.5 | 0 | 266 ± 39 |
| | | 166 | 2.5 | 23 | 1.3 | 142.9 | 209 ± 34 |
| | | 75 | 1.3 | 9 | 0.5 | 0 | 217 ± 28 |
| | green | 22 | 0.3 | 0 | 0 | 0 | 505 ± 89 |
| | | 22 | 0.3 | 0 | 0 | 0 | 422 ± 69 |
| | | 21 | 0.3 | 0 | 0 | 0 | 353 ± 47 |

TABLE I. Qualitative data for Polish power grid division into two, three and four zones. Each cell includes values for no wind / consensus / maximal wind generation.

## IV. CONCLUSIONS & FUTURE WORK

Our research shows how to extend the methodology of zones identification to take into account variable weather conditions. The approach is based on testing various weather scenarios and their aggregation by means of consensus clustering. We also point out two potential weaknesses of the method, i.e. (i) the problem of tiny clusters creation and (ii) the problem of intermingling zones, but many other issues related to stability and social welfare are standing in the queue.

Using the methodology described here we defined areas of similar energy cost which corresponds to the lack of congestion within these areas. Although this methodology is fully correct when defining zones, the price of energy in the area calculated by averaging nodal prices will not necessarily reflect the real "market" price anticipated in the zone. There are at least two reasons for that. First, when zonal market is considered, the market coupling algorithm is used instead of the exact OPF mechanism. Flow Based Market coupling assumes that zones are treated as copper plates connected by links characterized by PTDFs, which are then used to calculate the level of cross-border power exchange. The best situation is if these factors are insensitive for the distribution of the generation within a zone but depend only on aggregated generation in a zone. Otherwise transfers can be over- or underestimated and security limits on the branches can be potentially violated. Examination of PDTFs stability is naturally the next step of our further research. As the transfers found by market coupling are less accurate than those found by OPF, the market coupling prices can differ from those calculated by OPF.

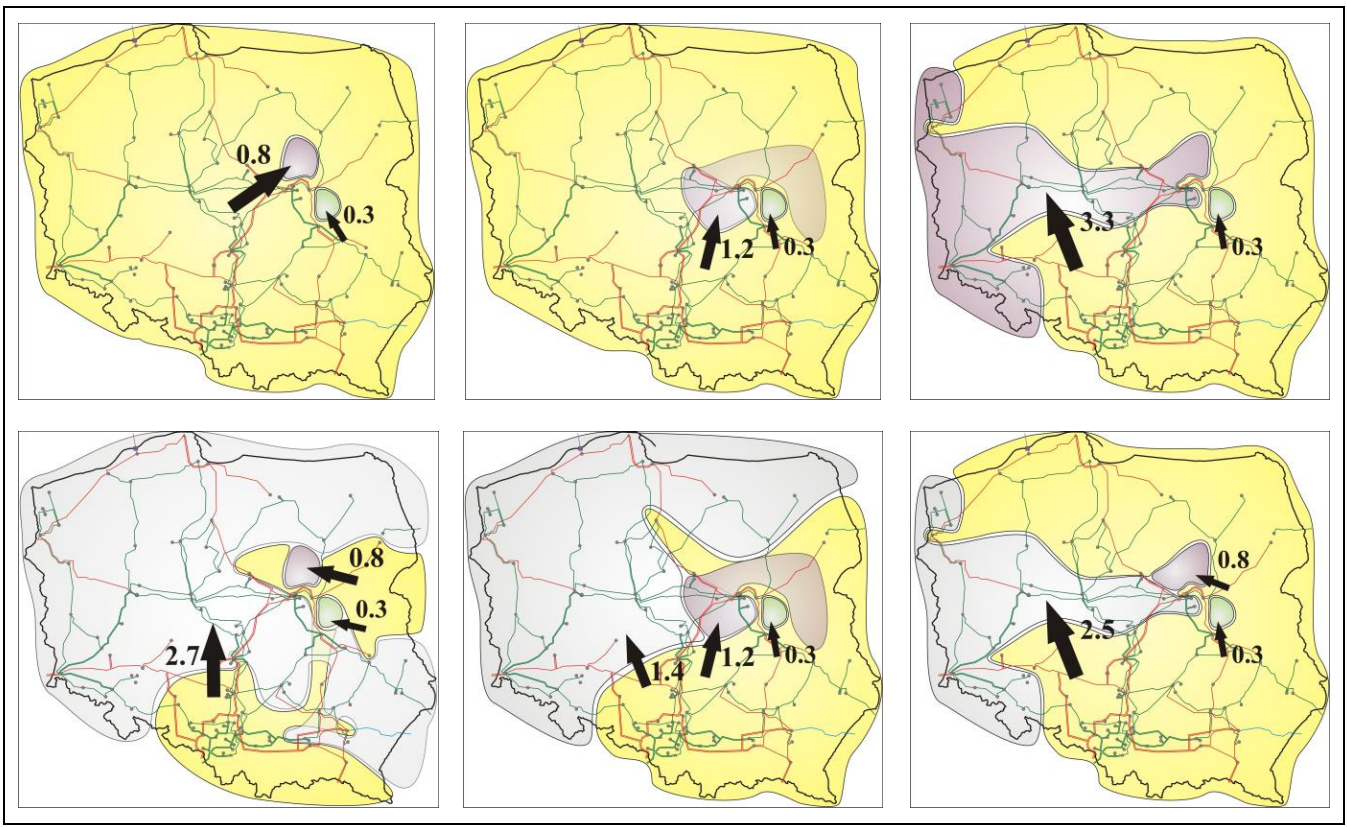

Figure 2. Polish power grid division for three (top) and four (bottom) zones. Results for no wind scenario, consensus clustering of wind cases and of maximal wind scenario are shown in the left, middle and right column, respectively. Consensus clustering reflects aggregating over 722 wind scenarios. Network scheme reflects extra high voltage lines, respectively, 400 kV (red), and 220 kV (green). Arrows indicate direction and magnitude of energy transfers between zones in GW.

Secondly, market participants (producers) can place bid offers which do not express their cost functions. The smaller the market zone is, the lower the number of market participants and the larger their impact on price. This creates the incentive for producers to conduct a strategic game. We intend to implement (i) market coupling and (ii) a model of market participants' strategic behavior in our future research.